\documentclass[12pt,english]{article}
\usepackage[dvipdfm]{color,graphicx}
\DeclareMathAlphabet{\mib}{OML}{cmm} {b}{it}
\definecolor{cyan}{cmyk}{1,0,0,0}
\definecolor{lightcyan}{cmyk}{0.5,0,0,0}
\definecolor{pastelcyan}{cmyk}{0.25,0,0,0}
\definecolor{magenta}{cmyk}{0,1,0,0}
\definecolor{yellow}{cmyk}{0,0,1,0}
\definecolor{lightyellow}{cmyk}{0,0,0.5,0}
\definecolor{pastelyellow}{cmyk}{0,0,0.25,0}
\definecolor{black}{cmyk}{0,0,0,1}
\definecolor{darkgray}{cmyk}{0,0,0,0.75}
\definecolor{gray}{cmyk}{0,0,0,0.5}
\definecolor{lightgray}{cmyk}{0,0,0,0.25}
\definecolor{white}{cmyk}{0,0,0,0}
\definecolor{red}{cmyk}{0,1,1,0}
\definecolor{orange}{cmyk}{0,0.5,1,0}
\definecolor{scarlet}{cmyk}{0,1,0.5,0}
\definecolor{brown}{cmyk}{0.5,0.75,1,0}
\definecolor{camel}{cmyk}{0.25,0.375,0.5,0}
\definecolor{cream}{cmyk}{0,0.2,0.3,0}
\definecolor{green}{cmyk}{1,0,1,0}
\definecolor{lightgreen}{cmyk}{0.5,0,0.5,0}
\definecolor{pastelgreen}{cmyk}{0.25,0,0.25,0}
\definecolor{mossgreen}{cmyk}{0.64,0.4,1,0}
\definecolor{yellowgreen}{cmyk}{0.5,0,1,0}
\definecolor{skyblue}{cmyk}{0.4,0.16,0,0}
\definecolor{royal}{cmyk}{1.0,0.5,0,0}
\definecolor{navyblue}{cmyk}{0.9,0.75,0.5,0}
\definecolor{blue}{cmyk}{1,1,0,0}
\definecolor{lightblue}{cmyk}{0.5,0.5,0,0}
\definecolor{lavender}{cmyk}{0.25,0.25,0,0}
\definecolor{violet}{cmyk}{0.75,1,0.25,0}
\definecolor{purple}{cmyk}{0.5,1,0.5,0}
\definecolor{pink}{cmyk}{0,0.5,0,0}
\definecolor{pastelpink}{cmyk}{0,0.25,0,0}
\def\e{{\rm e}}
\def\gtsim{\mathrel{\hbox{\raise0.2ex
\hbox{$>$}\kern-0.75em\raise-0.9ex\hbox{$\sim$}}}}
\def\ltsim{\mathrel{\hbox{\raise0.2ex
\hbox{$<$}\kern-0.75em\raise-0.9ex\hbox{$\sim$}}}}
\def\llt{\mathrel{<\kern-0.5em <}}   
\def\ggt{\mathrel{>\kern-0.5em >}}   

\def\Tr{{\rm Tr}}
\def\half{{1\over2}}
\def\kslash{k\kern-0.55em\raise 0.14ex\hbox{/}}
\def\Aslash{A\kern-0.6em\raise 0.14ex\hbox{/}}
\def\dslash{\del\kern-0.55em\raise 0.14ex\hbox{/}}
\def\therefore{ \hskip 0.0pt \raise0.1ex\hbox{.} \mkern -0.24mu \raise 1.1ex\hbox{.} \mkern -0.25mu \raise 0.1ex\hbox{.} \   } 
\def\because{ \hskip 0.0pt \raise 1.1ex\hbox{.} \mkern -0.24mu \raise 0.1ex\hbox{.} \mkern -0.25mu \raise 1.1ex\hbox{.} \  } 

\usepackage{babel}

\newcommand {\beq}{\begin{equation}}
\newcommand {\eeq}{\end{equation}}
\newcommand {\bea}{\begin{eqnarray}}
\newcommand {\eea}{\end{eqnarray}}
\newcommand {\nn}{\nonumber \\}

\newcommand {\m}{\mu}
\newcommand {\n}{\nu}
\newcommand {\pl}{\partial}
\newcommand {\p} {\phi}

\newcommand {\al}{\alpha}
\newcommand {\be}{\beta}

\newcommand {\La}{\Lambda}

\newcommand {\om}{\omega}
\newcommand {\Om}{\Omega}
\newcommand {\ep}{\epsilon}
\newcommand {\vep}{\varepsilon}
\newcommand {\na}{\nabla}
\newcommand {\del}  {\delta}
\newcommand {\Del}  {\Delta}
\newcommand {\mn}{{\mu\nu}}

\newcommand {\Ecal}{{\cal E}}

\newcommand {\Dcal}{{\cal D}}

\newcommand {\Wcal}{{\cal W}}

\newcommand {\phitil}{{\tilde \phi}}

\newcommand {\Ahat}{{\hat A}}   
\newcommand {\Bhat}{{\hat B}}   
\newcommand {\Dhat}{{\hat D}}   
\newcommand {\Ehat}{{\hat E}}   
\newcommand {\Hhat}{{\hat H}}

\newcommand {\vephat} {{\hat \varepsilon}}
\newcommand {\muhat} {{\hat \mu}}
\newcommand {\phihat} {{\hat \phi}}

\newcommand {\Ebar}  {{\bar E}}
\newcommand {\Bbar}  {{\bar B}}

\newcommand {\pdot}{\dot{p}}

\newcommand {\bx} {{\bf x}}
\newcommand {\bfZ} {{\bf Z}}
\newcommand {\A}{{\bf A}}
\newcommand {\B}{{\bf B}}
\newcommand {\D}{{\bf D}}
\newcommand {\E}{{\bf E}}
\newcommand {\bfH}{{\bf H}}
\newcommand {\K}{{\bf K}}

\newcommand {\bk}{{\bf k}}
\newcommand {\bq}{{\bf q}}
\newcommand {\bfna}{{\bf \nabla}}

\newcommand {\ABhat}{\hat {\bf A}}

\newcommand {\BBhat}{{\hat {\bf B}}}   
\newcommand {\DBhat}{{\hat {\bf D}}}   
\newcommand {\EBhat}{{\hat {\bf E}}}   
\newcommand {\HBhat}{{\hat {\bf H}}}   
\newcommand {\ABbar}{\bar {\bf A}}
\newcommand {\BBbar}{\bar {\bf B}}
\newcommand {\EBbar}{\bar {\bf E}}

\newcommand {\intfx} {{\int d^4x}}
\newcommand {\intcx} {{\int d^3x}}
\newcommand {\intck} {{\int d^3k}}

\newcommand {\intom} {{\int \frac{d\om}{2\pi}}}
\newcommand {\xperp}  {{{\bf x}_\perp   }}    

\newcommand {\ra} {\rightarrow}

\newcommand {\pr}   {{\quad .}}
\newcommand {\com}  {{\quad ,}}
\newcommand {\q}    {\quad}

\newcommand {\nl}    {\newline}

\newcommand {\PL}   {Phys.Lett.}

\newcommand {\PTP}  {Prog.Theor.Phys.}

\begin{document}
\title{
Renormalization Group Approach to Casimir Effect and 
the Attractive and Repulsive Forces in Substance 
\footnote{Proceedings of 
International Tribology Conference. Hiroshima 2011.10.30-11.03}}
\author{Shoichi Ichinose}
\maketitle

\begin{center}{
Laboratory of Physics, School of Food and Nutritional Sciences\\ 
University of Shizuoka, Yada 52-1, Shizuoka 422-8526, Japan\\
Corresponding author: ichinose@u-shizuoka-ken.ac.jp
                    }
\end{center}

\begin{abstract}
Electromagnetism in substance is characterized by 
permittivity (dielectric constant) and permeability (magnetic 
permeability). They describe the substance property {\it effectively}. 
We present a {\it geometric} approach to it. Some  
models are presented, where the two quantities are geometrically defined. 
Fluctuation due to the micro dynamics 
(such as dipole-dipole interaction) is taken into account 
by the (generalized) path-integral. 
Free energy formula (Lifshitz 1954), for the material composed of 
three regions with different permittivities, is explained. 
Casimir energy is obtained by a new regularization 
using the path-integral. Attractive force or repulsive one 
is determined by the sign of the {\it renormalization-group} $\beta$-function. 
\end{abstract}



\section{
{Introduction}
}
The recent development of the fundamental physics of the space-time-matter
(quantum field theory, string theory, D-brane theory) has revealed the interesting 
relation between the 'ordinary' physics (boundary field theory) and  the 'bulk space'
geometry dynamics. It is called AdS/CFT relation\cite{Malda9711,GKP9802,Witten9802}. 
One important outcome is that 
the renormalization-group flow can be regarded as the 'classical path' in the bulk 
(higher dimensional) 
space, which is called {\it holographic renormalization}\cite{SusWit9805, HenSken9806}. 
We interpret this new result in the context of the tribology physics. 

Casimir phenomena\cite{Casimir48} occur in substance at the (absolute) temperature zero:\ $T=0$. 
It is the zero-point oscillation of the {\it quantum} vacuum. Only {\it free}(kinetic) 
part contributes. 
Hence Casimir force (or energy) does {\it not} depend on couplings 
(interactions between the micro objects). 
The force works between two materials separated in 
the {\it macro} distance.  Besides, Casimir force sensitively depends on 
the {\it topology} of the material. 
To define Casimir energy rigorously we need highly-sophisticated regularizations because 
the quantity severely diverges both in the infrared(IR) and in the ultraviolet(UV) regions. 

In the science of friction, Casimir force (or energy) is one of important physical quantities 
where the renormalization procedure is necessary\cite{ICSF2010}. Usually the force works attractively. 
It plays the role of 
{\it friction}. The force (, for example, between two parallely-placed 
metallic plates) is experimentally observed and coincides with the theoretical result\cite{BMM01}.  
It is caused by the quantum fluctuation of the electromagnetic field. 
Another similar example is Van der Waals force. It is also caused by the {\it fluctuation} of 
the micro objects which compose the substance. The dipole-dipole interaction is 
an example of the fluctuation forces. 
Van der Waals force occurs at the general temperature $T$. The force also attractively works between
the neutral micro objects. The electromagnetism in substance describes this property, in terms of 
the dielectric constant (permittivity) $\vep$ and the magnetic permeability (permeability) $\m$, 
as the {\it effective} continuum theory. Van der Waals force 
reduces to Casimir one as $T\ra +0$ limit\cite{LL80}.

The electromagnetic field in substance is characterized by the relation between 
the electric field $\E$ (the magnetic field $\bfH$) and the electric flux density field $\D$ 
(the magnetic flux density field $\B$). 
\bea
\D=\vep(\om)\E\com\q \B=\mu(\om)\bfH\pr
\label{int1}
\eea 
\footnote
{The permittivity and the permeability will be definitely introduced later ( (\ref{ME2}), 
(\ref{ME7}), (\ref{ME11}) or (\ref{geo1}) ).
}
If the micro model of the material is given (such as the elastic force model for the electron 
in the atom), the forms of $\vep(\om)$ and $\mu(\om)$ are concretely obtained. Instead 
we approach the problem from the {\it geometrical} viewpoint. 
\section{
{Maxwell Equation in Substance }
}
  We consider the general continuous substance which has no 
{\it real} charge and current (classical vacuum) but has {\it induced} ones caused 
by the micro fluctuation. 
Let us explain the electromagnetism(EM) in substance with care 
for the $\om$(frequency), $t$(time) and $\bx$(space) dependence of 
the permittivity and the permiability. 
Electric and magnetic fields, $\EBhat(t,\bx)$ and $\HBhat(t,\bx)$, with their flux 
density fields, $\DBhat(t,\bx)$ and $\BBhat(t,\bx)$, are denoted as
\bea
\mbox{Upper-index Fields}\ :\q\q
\DBhat(t, \bx)=(\Dhat^i(x))\com\q \BBhat(t,\bx)=(\Bhat^i(x)), \nn
\mbox{Lower-index Fields}\ :\q\q 
\EBhat(t, \bx)=(\Ehat_i(x))\com\q \HBhat(t,\bx)=(\Hhat_i(x)), \nn
i=1,2,3,\ \bx=(x^i)=(x,y,z);\ \m=0,1,2,3,\ x=(x^\m)=(t,\bx).
\label{ME1}
\eea 
The dielectric function and the magnetic permeability are  
defined by (general form)
\bea
\Dhat^i(x)={ \vephat}^{ij}(x)\Ehat_j(x),\q\Bhat^i(x)=
\muhat^{ij}(x)\Hhat_j(x)
.
\label{ME2}
\eea 
We are usually considering in the 1+3 Minkowski (flat) space-time. The upper and lower 
indices appearing in (\ref{ME1}) indicate that some curved geometry 
is expected to describe the EM phenomena in substance {\it effectively}. We will   
fix the geometry later. 

The absence of real charges (electric and magnetic) requires the following conditions. 
\bea
\mbox{div}\DBhat =\pl_i\Dhat^i=0\q  \mbox{electric charge density}=0\com\nn
\mbox{div}\BBhat =\pl_i\Bhat^i=0\q  \mbox{magnetic charge density}=0
\com
\label{ME3}
\eea 
where $\pl_i\equiv \pl/\pl x^i$. 
Amp\`{e}re's and Faraday's laws are given by
\bea
\mbox{Amp\`{e}re's Law}\ :\q\q
\pl_t\Dhat^i-\ep^{ijk}\pl_j\Hhat_k=0\q\mbox{or}\q
\pl_t\DBhat-\bfna\times\HBhat=0,\nn 
(\mbox{electric current density}=0 ),\nn
\mbox{Faraday's Law}\ :\q\q
\pl_t\Bhat^i+\ep^{ijk}\pl_j\Ehat_k=0\q\mbox{or}\q
\pl_t\BBhat+\bfna\times\EBhat=0
,
\label{ME4}
\eea 
where $\pl_t\equiv \pl/\pl t$ and $\ep^{ijk}$ is the totally anti-symmetric 
tensor (Levi-Civita symbol) with $\ep^{123}=1$.

Faraday's law is solved by the vector and scalar potentials, $\ABhat(x)$ and $\phihat(x)
$.
\bea
\Ehat_i(x)=-\pl_t\Ahat_i(x)-\pl_i\phihat(x)
\ \mbox{or}\ \EBhat(x)=-\pl_t\ABhat(x)-\na\phihat(x)\nn
\Bhat^i(x)=\ep^{ijk}\pl_j\Ahat_k(x)\q\mbox{or}\q
\BBhat(x)=\na\times\ABhat(x)
,
\label{ME8}
\eea 
where $\ABhat(x)=\ABhat(t,\bx)=(\Ahat_i(x))$. 
Let us re-express above quantities in the Fourier-transformed form 
with respect to time $t$ (t-to-$\om$ Fourier-transformation).  
\bea
\DBhat(x)=\DBhat(t,\bx)=\int_{-\infty}^{\infty}\D(\om,\bx)\e^{i\om t}d\om\com\nn
\EBhat(x)=\EBhat(t,\bx)=\int_{-\infty}^{\infty}\E(\om,\bx)\e^{i\om t}d\om\com\nn
\BBhat(x)=\BBhat(t,\bx)=\int_{-\infty}^{\infty}\B(\om,\bx)\e^{i\om t}d\om\com\nn
\HBhat(x)=\HBhat(t,\bx)=\int_{-\infty}^{\infty}\bfH(\om,\bx)\e^{i\om t}d\om
\pr
\label{ME6}
\eea 
As for the permittivity and permeability, we consider 
the following form.
\bea
D^i(\om,\bx)=\vep^{ij}(\om)E_j(\om,\bx),\q
B^i(\om,\bx)=\mu^{ij}(\om)H_j(\om,\bx)
.
\label{ME7}
\eea 
\footnote{
Note that this form is {\it not} the most general one:\  
$\vep^{ij}=\vep^{ij}(\om,\bx)$ and $\m^{ij}=\m^{ij}(\om,\bx)$. 
          }
Note that we have replaced the definition of the permittivity and the permiability 
(\ref{ME2}) with the above one. 
Later we will consider some generalization of (\ref{ME7}). 

We also do t-to-$\om$ Fourier-transformation to the potentials.
\bea
\ABhat(t,\bx)=\int_{-\infty}^{\infty}\A(\om,\bx)\e^{i\om t}d\om,\q
\phihat(t,\bx)=\int_{-\infty}^{\infty}\phi(\om,\bx)\e^{i\om t}d\om
.
\label{ME9}
\eea 
The relations (\ref{ME8}) are re-expressed as
\bea
E_i(\om,\bx)=-i\om A_i(\om,\bx)-\pl_i\phi(\om,\bx)\ \mbox{or}\nn 
\E(\om,\bx)=-i\om\A(\om,\bx)-\bfna\phi(\om,\bx),\nn
B^i(\om,\bx)=\ep^{ijk}\pl_jA_k(\om,\bx)\q\mbox{or}\q
\B(\om,\bx)=\bfna\times\A(\om,\bx)
,
\label{ME9b}
\eea 
where $\bfna=(\pl_i)$.  
$\E(\om,\bx)$ and $\B(\om,\bx)$ are unchanged under 
the { gauge transformation}.                           
\bea
\A\q\ra\q \A+\bfna\La,\q \phi\q\ra\q \phi - i\om\La
.
\label{ME10}
\eea 
where $ \La=\La(\om,\bx)$ is the local gauge freedom. 

For simplicity, we consider the diagonal permittivity and permeability. 
\bea
\vep^{ij}={ \vep}(\om)\del^{ij}\com\q  ({ \mu^{-1}})_{ij}={ \mu}^{-1}(\om)\del_{ij} 
\com
\label{ME11}
\eea 
where $(\del^{ij})=(\del_{ij})=\mbox{diag}(1,1,1)$.\nl
                             
{\bf Gauge 1}
\nl
\ First we take the following gauge.
\bea
\pl_i\{i\om\phi + (\vep\mu)^{-1}\mbox{div}\A\}=0
.
\label{ME11b}
\eea 
Amp\`{e}re's law gives the field eq. of $\A$ 
\bea
(\Del+\om^2\vep\mu)\A(\om,\bx)=0,\q 
\E=-i\om\A-\frac{i}{\om\vep\mu}\bfna(\mbox{div}\A)
.
\label{ME12}
\eea 
When $\vep$ and $\m$ are constants (do not depend on $\om$), $\vep=\vep_1, \m=\m_1$,\ 
$\ABhat(x)$ satisfies the {\it free} wave equation with the velocity $v=1/\sqrt{\vep_1\m_1}$. 
\bea
(\Del-\frac{1}{v^2}\frac{\pl^2}{\pl t^2})\ABhat(t,\bx)=0,\nn 
v=1/\sqrt{\vep_1\m_1},\q
\ABhat(t,\bx)=\int\A(\om,\bx)\e^{i\om t}d\om
.
\label{ME12b}
\eea 
We keep the case:\ $\vep=\vep(\om),\mu=\mu(\om)$. 
From the condition $\mbox{div}\D=\vep~\mbox{div}\E=0$, which is derived from (\ref{ME3}), 
we obtain
\bea
(\Del+\om^2\vep\mu)\mbox{div}\A(\om,\bx)=0
\pr
\label{ME12c}
\eea 
Using this equation, we can show the energy density $\Ecal$ is given by
\bea
\Ecal=\half (\E \cdot \D+\bfH \cdot \B)
=\half (\vep^{ij} E_i E_j+{\m^{-1}}_{ij} B^i B^j)\nn
=\half \m^{-1}\A \cdot (\Del+\om^2\vep\m)\A +\mbox{total derivative}
\pr
\label{ME12d}
\eea 
\nl

{\bf Gauge 2}
\footnote{
This gauge is used in the Landau-Lifshitz textbook\cite{LL80}
          }
\nl
\ \ We can take another gauge.
\bea
\pl_i\phi=0
\pr
\label{ME12e}
\eea 
Amp\`{e}re's law (\ref{ME4}) gives the field eq. of $\A$
\bea
\Del\A-\bfna(\mbox{div}\A)+\om^2\vep\mu\A=0,\q
\E=-i\om\A
.
\label{ME13}
\eea 
From the condition $\mbox{div}\D=\vep~\mbox{div}\E=0$, we obtain
\bea
\mbox{div}\A=-\frac{1}{i\om}\mbox{div}\E=0
\pr
\label{ME14}
\eea 
Hence the field eq. (\ref{ME13}) reduces to (\ref{ME12})
in the present case of {\it no charge and no currents}(classical vacuum). 
In this gauge too, we can confirm the relation (\ref{ME12d}).

\section{
Geometry in 4D space (($K^\m$)=($\om$,$K^i$)) and 
Induced Geometry in 3D space (($k^i$)) \label{geo}
}

\begin{figure}
\caption{
3 dim hyper-surface (\ref{geo3}). $0\leq\om\leq T$. Renormalization group flow. 
        }
\begin{center}
\includegraphics[height=7cm]{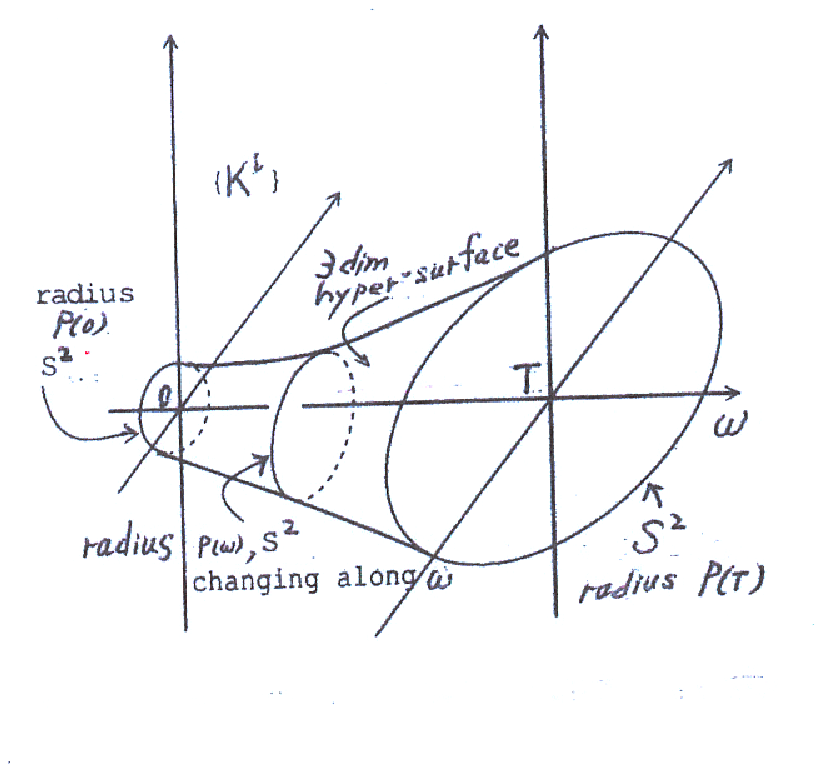}
\end{center}
\label{RefFlowSurf}
\end{figure}

From the previous description, we know the energy can be 
expressed by either $\A(\om,\bx),\E(\om,\bx)$ and $\B(\om,\bx)$, 
\bea
H'=\intcx\int d\om~\Ecal
=\intcx\int d\om\half (\vep^{ij} E_i E_j+{\m^{-1}}_{ij} B^i B^j)\nn
=\intcx\int d\om\half \m^{-1}\A \cdot (\Del+\om^2\vep\m)\A
\com
\label{geo1}
\eea 
or 
their $\bx$-to-$\bk$ Fourier-transformed ones $\ABbar(\om,\bk),\EBbar(\om,\bk)$ 
and $\BBbar(\om,\bk)$.
\bea
H'
=\intck\int d\om\half (\vep^{ij}(\om,\bk) \Ebar_i(\om,\bk) \Ebar_j(\om,\bk)
+{\m^{-1}(\om,\bk)}_{ij} \Bbar^i(\om,\bk) \Bbar^j(\om,\bk))\nn
=\intck\int d\om\half \m^{-1}(\om,\bk)\ABbar(\om,\bk)\cdot (-\bk^2+\om^2\vep(\om,\bk)\m(\om,\bk))
\ABbar(\om,\bk)
\pr
\label{geo1b}
\eea 
We stress again that the forms of $\vep^{ij}$ and $\mu^{ij}$ 
represent the substance property. We now specify the forms in the geometrical way. 
\footnote{
The metric treatment of the permittivity and the permeability has been frequently 
suggested in the past\cite{Tamm24, Skrotskii57, Plebanski60, GHWW09}. 
          }
First let us {\it introduce} the metric $G_\mn(K)$ in the 4 dim space $(\om,K^i)\equiv (K^\m)$, 
$ds^2=G_\mn(K)dK^\m dK^\n$. 
As interesting metrics, we can consider the following ones.  
\bea
\begin{array}{ll}
\mbox{1. Minkowski}   &   ds^2=-d\om^2+\sum_{i=1}^3{dK^i}^2  \\
\mbox{2. dS}_4          &   ds^2=-d\om^2+\e^{2H_0\om}\sum_{i=1}^3{dK^i}^2\com\q H_0>0\\
\mbox{3. AdS}_4        &   ds^2=(dK^3)^2+\e^{-2H_0|K^3|}(-d\om^2+(dK^1)^2+(dK^2)^2)
\end{array}
\label{geo2}
\eea 
\footnote{
Three types of metric, \ 
1) Minkowski, 2) de Sitter (dS), 3) anti de Sitter (AdS),  are all {\it maximally symmetric}. 
$H_0$ is a model parameter ( constant ) which expresses the 4 dim curvature. 
          }
$H_0$ is a model parameter (constant) which expresses the 4 dim curvature. 
With the aim of specifying (parametrizing) the 3 dim metric $g_{ij}(\om)$, we {\it introduce} 
3 dim {\it hyper-surface} in this 4 dim space $(\om,K^i)$. 
\bea
\mbox{Dispersion relation}:\q (k^i)^2\ =\ p(\om)^2
,\label{geo3}
\eea 
where the {\it isotropy} of the 3 spacial directions ($K^1,K^2,K^3$) 
is assumed, and $p(\om)$, which specify the hypersurface,  
is some function (of $\om$) to be explained below. 
See Fig.\ref{RefFlowSurf}. 
\footnote{
Fig.\ref{RefFlowSurf} shows the renormalization flow in the "holographic approach". 
We can regard the 3 dim hyper-surface as the "physical world" where the actual 
(not formally artificial) things occur. We are looking the physical world from the 
higher-dimensional (4 dim) space. On the point $\om$ of the $\om$-axis 
(called "extra-axis"), there exists 3 dim {\it ball} with the boundary of $S^2$ sphere (radius p($\om$)). 
The center of the ball is $(\om, \K={\bf 0})$. 
The ball is like a {\it brane}. If we can regard the radius $p(\om)$ as a scale, 
the whole configuration surrounded by the hyper-surface describes 
the scale flow of the system. The "brane" moves along the $\om$-axis changing its radius $p(\om)$. 
}
In the following, we explain taking the case 1, Minkowski metric. 

\begin{figure}
\caption{
Behavior of $p_1(\om)$ in (\ref{geo5}). Dispersion relation of the massless particle. 
        }
\begin{center}
\includegraphics[height=7cm]{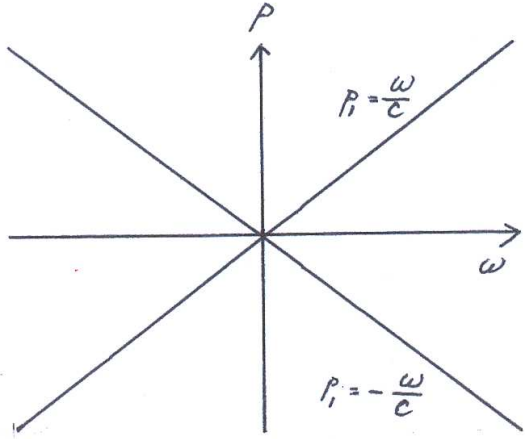}
\end{center}
\label{MasslessDisp}
\end{figure}

\begin{figure}
\caption{
Behavior of $p_2(\om)$ in (\ref{geo5}). Dispersion relation of the massive particle.
        }
\begin{center}
\includegraphics[height=7cm]{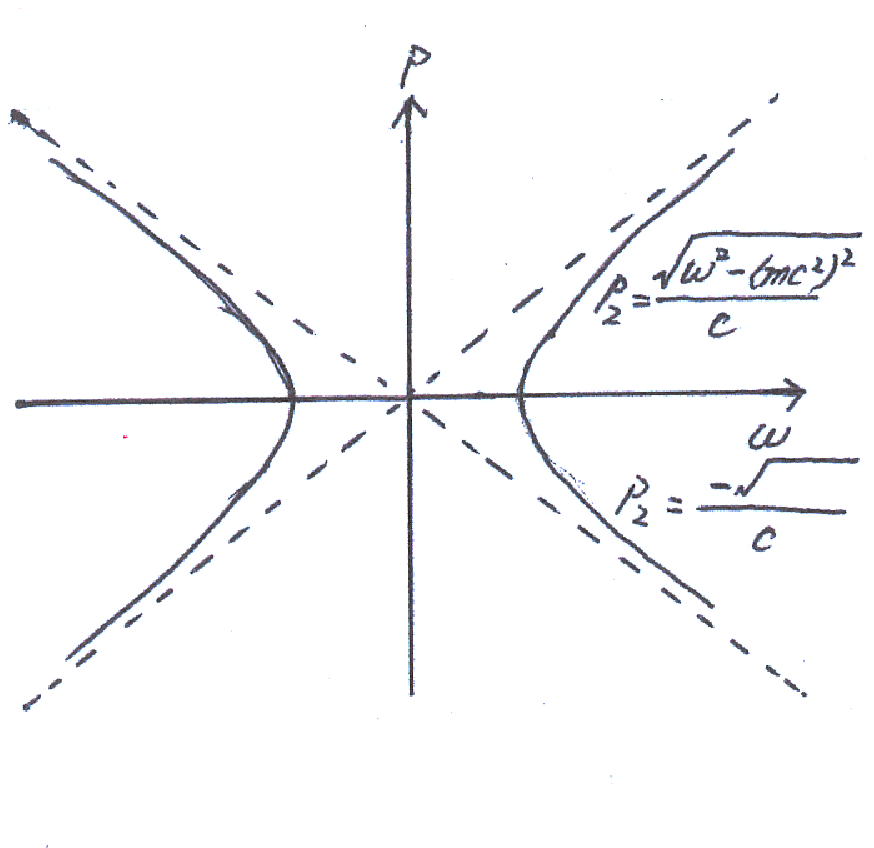}
\end{center}
\label{MassiveDisp}
\end{figure}

The {\it induced} metric $g_{ij}$, is given as
\bea
ds^2=\{ -(p\pdot)^{-2}k^ik^j+\del_{ij} \}dk^idk^j=g_{ij}(\om,\bk)dk^idk^j\com\q\pdot=\frac{dp}{d\om}\ ,
\label{geo4}
\eea 
When $p(\om)$ is specified, $g_{ij}(\om,\bk)$ is explicitly given. 
We will soon show how to geometrically determine the form
\footnote{
See eq.(\ref{geo12}). 
}
, but 
we here show some examples. 
\bea
g_{ij}(\om,\bk)=\left\{ \begin{array}{ll}
  \del_{ij}-\frac{c^4}{\om^2}k_{(1)}^ik_{(1)}^j,    & 
\begin{array}{c}
p_1(\om)=\frac{\om}{c},\ \pdot_1=\frac{1}{c}, \\
  (k_{(1)}^i)^2=\frac{\om^2}{c^2}
\end{array}
                                                                      \\
  \del_{ij}-\frac{c^4}{\om^2}k_{(2)}^ik_{(2)}^j,    &
\begin{array}{c}
p_2(\om)=\frac{\sqrt{\om^2-(mc^2)^2}}{c},\\
\pdot_2=\frac{\om}{c^2}\frac{1}{p_2}, (k_{(2)}^i)^2=\frac{\om^2-(mc^2)^2}{c^2}
\end{array}
\end{array}
                  \right.
\com\label{geo5}
\eea 
where $c$(light velocity) and $m$(mass) are some constants. See Fig.\ref{MasslessDisp} and Fig.\ref{MassiveDisp}. 

The system energy expression (\ref{geo1b}) is the summation over all frequency-momentum
points ($\om,\bk$). The quantity, however, is divergent. Let us {\it propose} that the summation 
should be replaced by that of all hypersurfaces (\ref{geo3}). 
From the requirement of the {\it general coordinate invariance}, the energy 
expression $H'$ of (\ref{geo1b}) is replaced by 
\bea
H''=
\int d[\mbox{hyper-surface(\ref{geo3})}]\times\nn
\half \sqrt{\mbox{det}g_{ij}(\om,\bk)}
(\vep^{ij}(\om,\bk) \Ebar_i(\om,\bk) \Ebar_j(\om,\bk) \nn 
+{\m^{-1}(\om,\bk)}_{ij} \Bbar^i(\om,\bk) \Bbar^j(\om,\bk))\nn
\equiv\int d[\mbox{hyper-surface(\ref{geo3})}]\sqrt{\mbox{det}g_{ij}(\om,\bk)}{\bar \Ecal}[\ABbar(\om,\bk)]
\com
\label{geo6}
\eea 
where we assume 
\bea
\vep^{ij}(\om,\bk) = e_1 g^{ij}(\om,\bk),\q
\m(\om,\bk)^{ij}=m_1 g^{ij}(\om,\bk)
,
\label{geo7}
\eea 
where $e_1$ and $m_1$ are constants. 
Note that the expression $H''$ can be written by the vector potential 
$\A(\om,\bx)$ in the non-local form. 

We notice that $H''$ depends on the hyper-surface (\ref{geo3}). 
There are many hyper-surfaces by {\it varying} the form of $p(\om)$. 
The present {\it model} of the electromagnetism in substance should 
describe the {\it fluctuation} of the micro dynamics. 
In order to take it into account, we {\it propose} here 
to promote the expression $H''$, 
(\ref{geo6}), to the following {\it generalized} path-integral expression (\ref{geo10})
\footnote{
The ordinary path-integral is the summation over all possible lines (under the given boundary 
condition) , whereas the present one is over all possible hyper-surfaces. 
}.

Before presenting (\ref{geo10}), we explain a geometrical quantity, {\it area} $A$, of 
the hyper-surface. 
On the hyper-surface (\ref{geo3}),\  
$\sum_{i=1}^3(k^i)^2=p(\om)^2$, the {\it induced metric} $g_{ij}$
gives us the {\it area} as the functional of the path $\{ p(\om): 0\leq\om\leq T\} $ where 
$T$ is introduced as the upper bound (a boundary parameter) for the frequency $\om$. 
\bea
A[p(\om)]=\int\sqrt{\det g_{ij}}~d^3k=\int_0^T \sqrt{\pdot^2+1}~p^2d\om
.
\label{geo8}
\eea 
In order to express the {\it statistical ensemble} due to the micro fluctuation
we take the following {\it distribution} $\Om[p(\om)]$ for the energy expression (\ref{geo6}). 
\bea
\Om[p(\om)]= \frac{1}{N}\exp(-\frac{1}{2\al'}A[p(\om)])\nn
=\frac{1}{N}\exp\{
-\frac{1}{2\al'}\int_0^T \sqrt{\pdot^2+1}~p^2d\om
                \}
,
\label{geo9}
\eea 
where a {\it new} model parameter $\al'$ ({\it string tension}) is introduced. $N$ is the normalization factor. Taking the above distribution $\Om$, 
the system energy $H=E(T)$ is finally given by  
\bea
E(T)=\frac{1}{N}\int_{0}^{\infty}d\rho
\int_{\begin{array}{c}p(0)=\rho\\p(T)=\rho\end{array}}
\prod_{\om,i}\Dcal k^i(\om)\times\nn 
{\bar \Ecal}[\ABbar(\om,\bk)]
\exp \left[
-\frac{1}{2\al'}\int_0^T \sqrt{\pdot^2+1}~p^{2}d\om
                                  \right],
\label{geo10}
\eea 
where the integral is over all hyper-surfaces (\ref{geo3}) and 
${\bar \Ecal}[\ABbar(\om,\bk)]$ is defined in (\ref{geo6}). 

Among all possible paths $\{ p(\om): 0\leq\om\leq T\}$, the dominant one is given by the minimal 
area principle.
\bea
\del A[p(\om)]=\del \int_0^T \sqrt{\pdot^2+1}~p^2d\om =0
\pr
\label{geo12}
\eea 
The solution is explained in ref.\cite{SI0801}. In the calculation of (\ref{geo10}), 
infrared (IR) and ultraviolet (UV) divergences appear. To regularize them, we calculate 
with the IR-cutoff $\mu$($\ra +0$) and UV-cutoff $\La$($\ra +\infty$).
\bea
\int_0^\infty d\rho\ \cdots\q \ra\q \int_\mu^\La d\rho\ \cdots 
\pr
\label{geo13}
\eea 
Log-divergence ($\ln \La$) appears\cite{SI0801} and it is renormalized by the 
{\it boundary} parameter $T$. This is the case of Minkowski. For other cases 
(de Sitter, anti de Sitter), the parameter $H_0$ is renormalized\cite{SI0812,SI0909,SI1205}.

\begin{figure}
\caption{
Configuration of the material composed of three regions with permittivity $\vep_1, \vep_m, \vep_2$.  
        }
\begin{center}
\includegraphics[height=7cm]{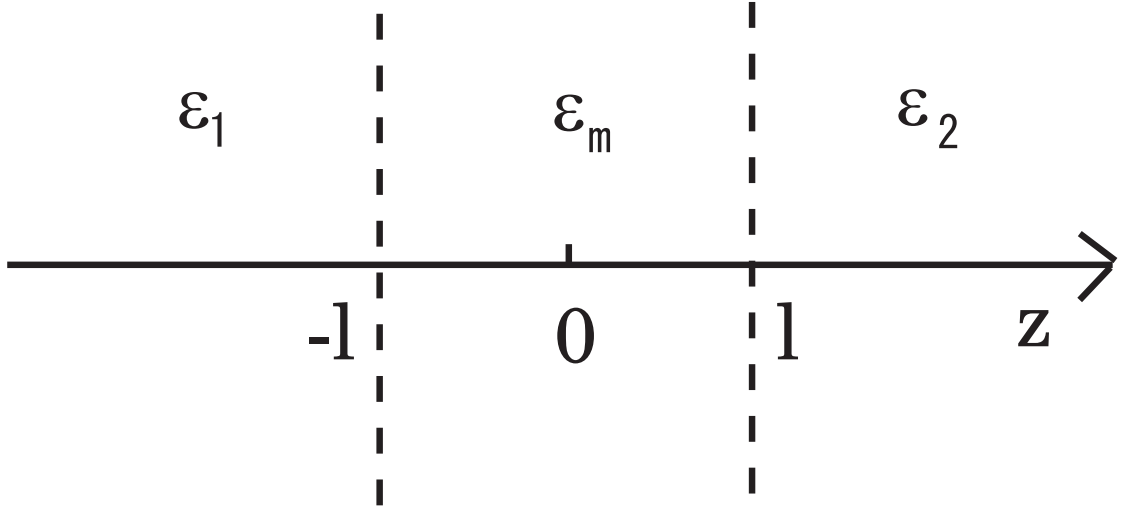}
\end{center}
\label{ICTfig1}
\end{figure}

\section{
Lifshitz Formula
}

Let us take the material composed of three regions with 
different permittivities $\vep_1(\om),\vep_m(\om),\vep_2(\om)$ ($\mu=1$) , 
(Fig.\ref{ICTfig1}). 
\footnote{
$\vep(\bx,\om)=\{\vep_1(\om)\ \mbox{when}\ \bx\in R_1,\ \ \vep_m(\om)\ \mbox{when}\ \bx\in R_m, 
\ \ \vep_2(\om)\ \mbox{when}\ \bx\in R_2\}$
}
We consider 
the free energy  at temperature $T$. 
Instead of the action derived from (\ref{ME12}), we take 
the following simplified model of the Maxwell theory\cite{KK06}. 
\bea
S=\half\intcx\intom \phi^*_\om (\Del+\om^2\vep(\om))\phi_\om,\nn
\phi^*_\om=\phi_{-\om},\q \vep(\om)=1+\chi(\om)
.
\label{LT1}
\eea 
This is the action for each region. 
Compare the simplified model above with that of Maxwell theory (\ref{geo1}). 
When $\vep(\om)=c_1$(const.), the above expression is the action of the (3+1) dim massless complex 
{\it free} scalar.
\bea
S_{free}=\half\intcx\int\frac{dt}{2\pi} {\hat \phi}^*(\bx,t) (\Del-c_1\frac{\pl^2}{\pl t^2}){\hat \phi}(\bx,t),\nn
{\hat \phi}(\bx,t)=\int_{-\infty}^{\infty}\phi_\om(\bx)\e^{i\om t}d\om
.
\label{LT2}
\eea 
We keep the general case (\ref{LT1}). The field equation is given by 
\bea
(\Del+\om^2\vep(\om))\phi_\om=0,\q \phi_\om(\xperp,z)=\phitil_\om(z)\e^{i\bq\cdot\xperp}
.
\label{LT3}
\eea 

This system is in the thermal equilibrium at temperature $T$. 
We can realize this situation by imposing the {\it periodicity} condition on the time variable $t$. 
The period is $1/T$. 
\bea
\mbox{Periodicity}:\q t\ \ra\ t\ +\ \frac{1}{T},\q
\om_n=\frac{2\pi T}{\hbar}n
.
\label{LT4}
\eea 
In the plane perpendicular to z-axis, we impose 
the periodicity with the length $L$. This is 
for the IR regularization. $L$ is considered 
to be sufficiently large. 
\bea
\mbox{Periodicity}:\q \xperp\equiv (x,y)\ \ra\ (x+L,y+L),\nn
\bq_{(n_x,n_y)}=(\frac{2\pi}{L}n_x, \frac{2\pi}{L}n_y)
.
\label{LT5}
\eea 
The wave function $\phitil_\om(z)$ in (\ref{LT3})
satisfies. 
\bea
(-\bq^2+{\pl_z}^2+\om^2\vep(\om))\phitil_\om(z)=0
\pr
\label{LT6}
\eea 

Assuming the form $\phitil_\om(z)\propto \e^{\pm \rho z}$, we obtain 
\bea
\phitil^j_\om(z)=A_j(\om)\e^{\rho_j z}+B_j(\om)\e^{-\rho_j z},\nn
-\bq^2+{\rho_j}^2+\om^2\vep_j(\om)=0\q (j=1,m,2)
.
\label{LT6b}
\eea 
The proper boundary condition (damping in the remote regions)  
finally determines the wave function for each region as
\bea
\begin{array}{ll}
\mbox{region 1}  &
z<-l,\q  \phitil_\om(z)=A(\om)\e^{\rho_1z},\\
\mbox{region m} &
-l<z<l,\q \phitil_\om(z)=C_1(\om)\e^{\rho_mz}+C_2(\om)\e^{-\rho_mz},\\
\mbox{region 2}  &
z>l,\q  \phitil_\om(z)=B(\om)\e^{-\rho_2z}
\end{array}
\label{LT7}
\eea 
Imposing the {\it continuity} and the {\it smoothness} at the two boundaries, 
the boundary between $R_1$ and $R_m$ and that between $R_m$ and $R_2$, 
we can get the solution under the condition:
\bea
\Del=1-\frac{(\rho_1-\rho_m)(\rho_2-\rho_m)}{(\rho_1+\rho_m)(\rho_2+\rho_m)}\e^{-4\rho_ml}=0
\pr
\label{LT8}
\eea 
This condition comes from avoiding the trivial solution: $A=B=C_1=C_2=0$. 

Now we can define the free energy $F$ using the path-integral. 
\bea
\e^{-F_\chi}=\int\Dcal\phi_\om\Dcal\phi^*_\om \e^{iS[\phi^*,\phi;\chi_1,\chi_m,\chi_2]}\nn
=\det (\Del+\om^2\vep_\al(\om))
=\exp \Tr \ln~(\Del+\om^2(1+\chi_\al(\om))~)
\com
\label{LT10}
\eea 
where $\vep_\al=1+\chi_\al$ is given by
\bea
\mbox{In R}_1\q 1+\chi_1(\om)=\frac{1}{\om^2}(\frac{2\pi}{L})^2(n_x^2+n_y^2)
                                           -\frac{{\rho_1}^2}{\om^2}\com\nn
\mbox{In R}_m\q 1+\chi_m(\om)=\frac{1}{\om^2}(\frac{2\pi}{L})^2(n_x^2+n_y^2)
                                           -\frac{{\rho_m}^2}{\om^2}\com\nn
\mbox{In R}_2\q 1+\chi_2(\om)=\frac{1}{\om^2}(\frac{2\pi}{L})^2(n_x^2+n_y^2)
                                           -\frac{{\rho_2}^2}{\om^2}
\com
\label{LT10b}
\eea 
If the three functions $\chi_1(\om),\chi_m(\om),\chi_2(\om)$, which characterize 
the material, are given, the wave function is completely solved at the 
IR-regularized (xy-plane) level. The free energy defined in (\ref{LT10}) still has 
UV-divergences. To deal with it, we use the ambiguity of energy origin $F=0$.\nl  
\nl
REGULARIZATION 1\nl
First we may subtract the value at $\chi=0$. This is because we suppose 
the region of no substance does not contribute to the system energy. 
\bea
F_C\equiv F_\chi -F_{\chi=0}=-\Tr \ln (1+\frac{\om^2\chi_\al(\om)}{\Del+\om^2})
\pr
\label{LT11}
\eea 
Next we may subtract trivial constants. We have interest only in the 
interaction between different materials, not in the each material's property. \nl
\nl
REGULARIZATION 2 (Entanglement)
\bea
F\equiv F_C(R_1\cup R_m\cup R_2)-F_C(R_1)-F_C(R_m)-F_C(R_2)
.
\label{LT12}
\eea 
Finally we obtain the well-defined (finite) quantity\cite{KK06}. 

In Ref.\cite{KK06}, it is shown that the Casimir force between two materials related by 
{\it reflection} is always {\it attractive}. 


\begin{figure}
\caption{
Configuration of Casimir energy measurement. 
        }
\begin{center}
\includegraphics[height=8cm]{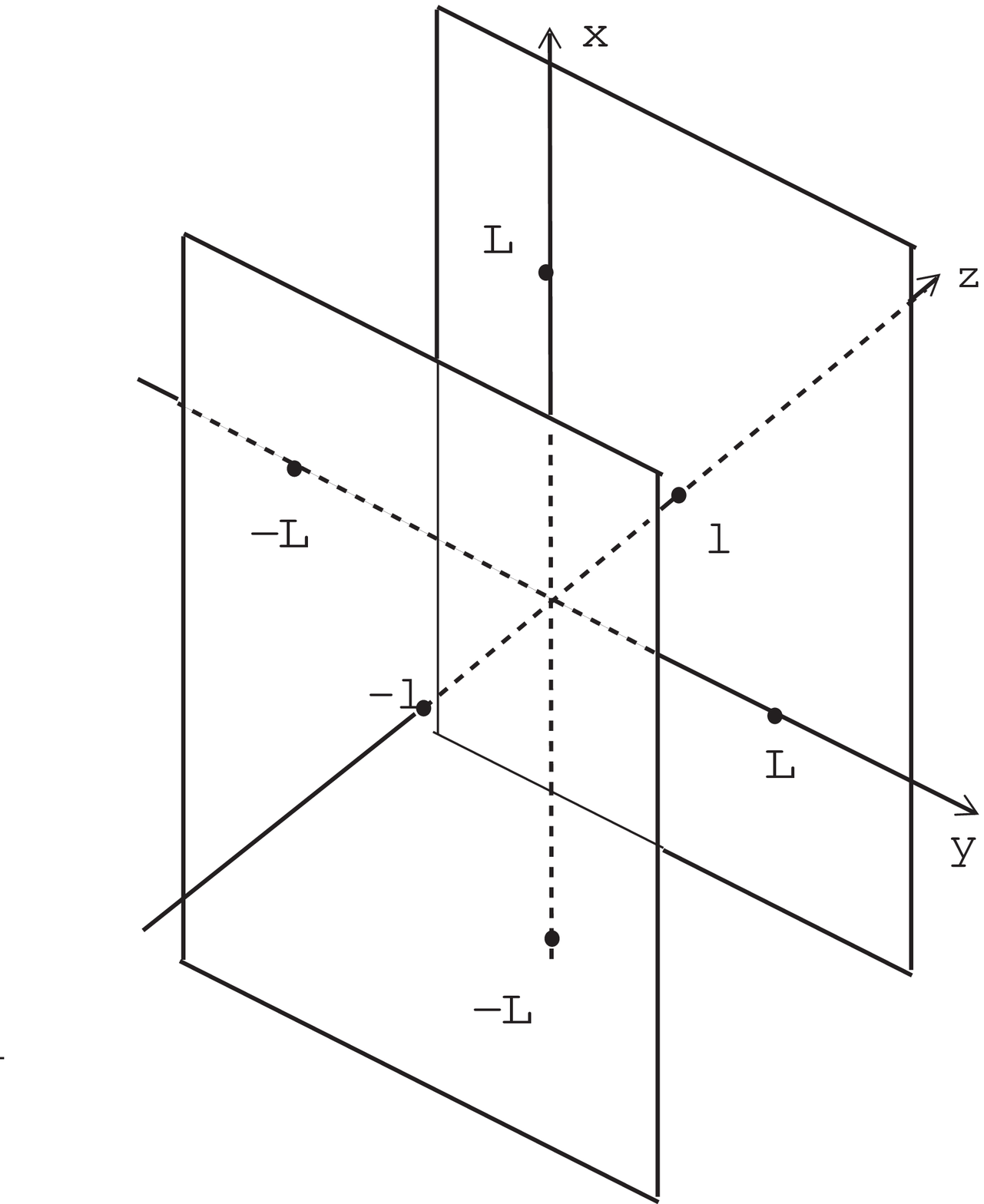}
\end{center}
\label{2Plates}
\end{figure}
\section{
Ordinary Regularization for Casimir Energy\label{CE}
}
Let us consider 
1+3 dim electromagnetism (free field theory, $(\vep_0, \m_0)$: constants of vacuum values) 
in Minkwski space (\ref{ME13}), (\ref{ME14}): 
\bea
S=\intfx\half\A\cdot(\Del-\frac{1}{c^2}\frac{\pl^2}{\pl t^2})\A=0,\q c
=\frac{1}{\sqrt{\m_0\vep_0}}\equiv 1,\nn
ds^2=-dt^2+dx^2+dy^2+dz^2
.
\label{CE1}
\eea 
2 perfectly-conducting plates parallel with the separation $2l$ 
in the z-direction. See Fig.\ref{2Plates}.   
As for x- and y-directions, we impose the periodicity $2{ L}$ for the { IR regularization}.
\bea
\mbox{Periodicity} :\q
x\ra x+2L,\q
y\ra y+2L,\q
z\ra z+2l
,\nn
L\gg l
\com
\label{CE2}
\eea
The eigen frequencies and Casimir energy are 
\bea
\om_{m_x,m_y,n}=\sqrt{(n\frac{\pi}{l})^2+(m_x\frac{\pi}{L})^2+
(m_y\frac{\pi}{L})^2}\com\nn
E_{Cas}=2\cdot\sum_{m_x,m_y,n\in \bfZ}\half\om_{m_x,m_y,n}\q \geq 0
\com
\label{CE3}
\eea
where $\bfZ$ is the set of all integers.  
$\half\om_{m_x,m_y,n}$ is the { zero-point oscillation energy}. 

Introducing the cut-off function $g(x)$ 
(= 1 [$0<x<1$] or 0 [otherwise]),  
\bea
E_{Cas}^{\La}=\sum_{m_x,m_y,n\in \bfZ}\om_{m_x,m_y,n}g\left(\frac{\om_{m_x,m_y,n}}{\La}\right)
\q\geq 0
,
\label{CE4}
\eea
where $\La$ is the UV-CutOff. 
Taking the continuum limit $L\ra\infty,\ L\ll l\ra\infty$, we obtain 
\bea
E_{Cas}^{\La 0}=\int_{-\infty}^{\infty}\int_{-\infty}^{\infty}\frac{dk_xdk_y}{(\frac{\pi}{L})^2}
\int_{-\infty}^{\infty}\frac{dk_z}{\frac{\pi}{l}}\sqrt{k_x^2+k_y^2+k_z^2}~g(\frac{k}{\La})\nn
={\int_{-\infty}^{\infty}\int_{-\infty}^{\infty}\int_{-\infty}^{\infty}}_{|k|\leq\La}
\frac{dk_xdk_ydk_z}{(\frac{\pi}{L})^2\frac{\pi}{l}}
\sqrt{k_x^2+k_y^2+k_z^2}\q\geq 0
.
\label{CE5}
\eea
Note that $E_{Cas}$, $E_{Cas}^\La$ and $E_{Cas}^{\La 0}$ 
are all { positive-definite}. 
In a familiar way, regarding $E_{Cas}^{\La 0}$ as 
the { origin of the energy scale}, we consider 
the quantity $u=(E_{Cas}^{\La}-E_{Cas}^{\La 0})/(2L)^2$ 
as the physical Casimir energy and evaluate it with the help 
of the { Euler-MacLaurin formula} as $u=(\pi^2/(2l)^3)~({ B_4}/4!)=-(\pi^2/720)(1/(2l)^3)<0$.
\footnote{
$B_4$ is the 4-th Bernoulli number. 
} 
The final result is { negative}. In the present analysis we take a { new} regularization which { keeps positive-definiteness}.

\section{
New Regularization for Casimir Energy
}
\begin{figure}[h]
\begin{minipage}{25pc}
\includegraphics[width=25pc]{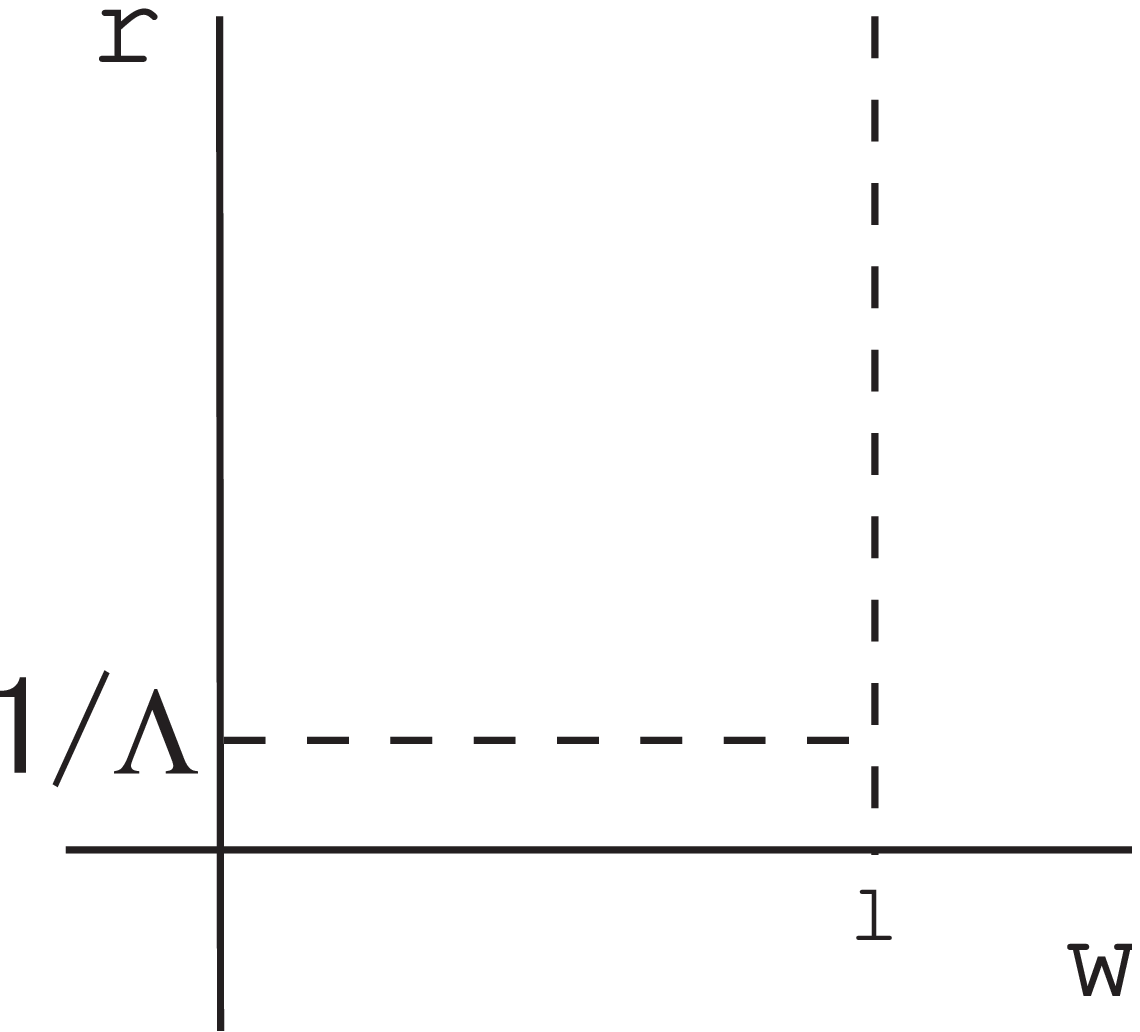}
\caption{\label{2DregionI9}The integral region of (\ref{CE6}). 
}
\end{minipage}\hspace{2pc}%
\begin{minipage}{25pc}
\includegraphics[width=25pc]{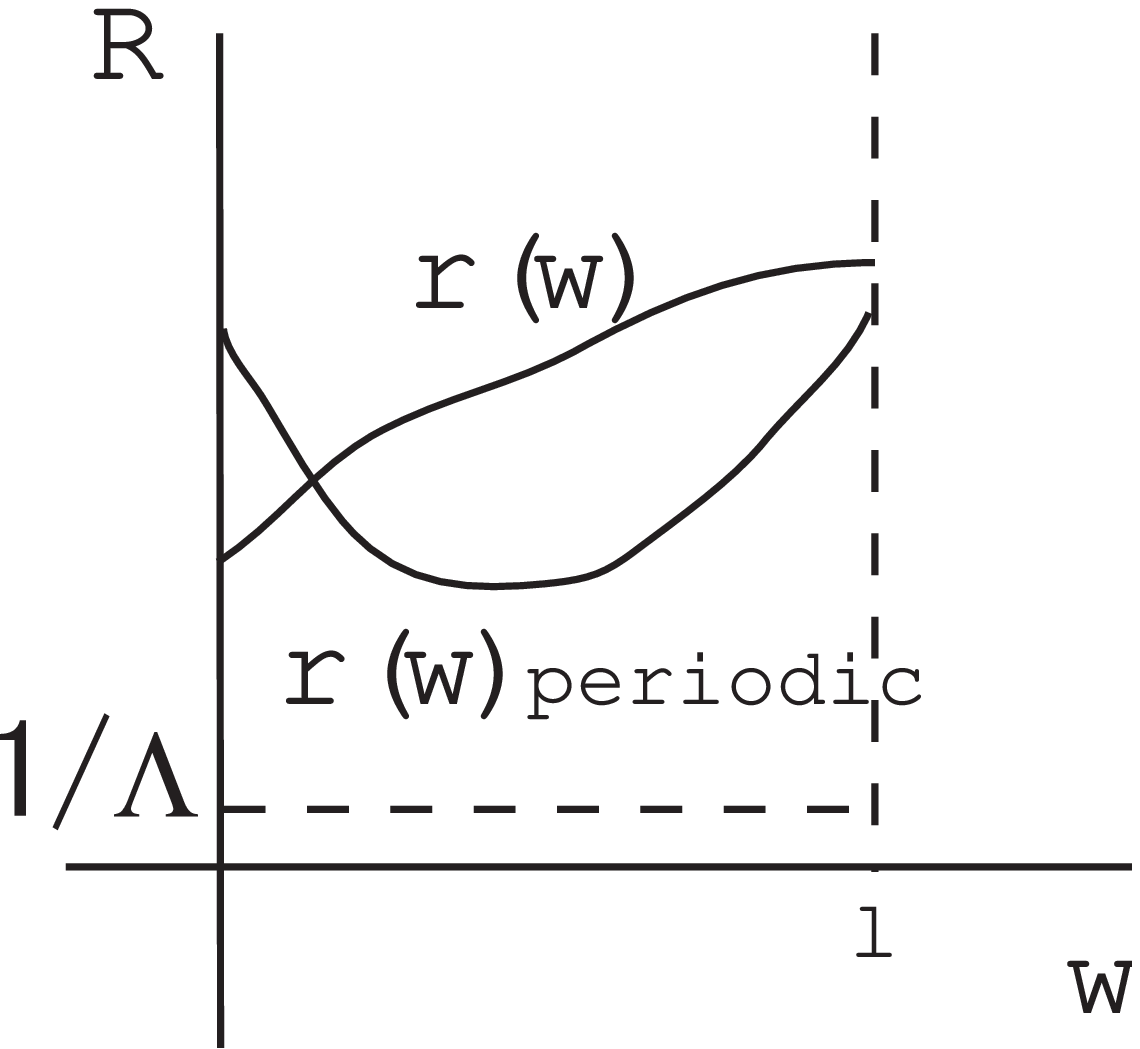}
\caption{\label{2DpathI9}A general path $r(w)$ of (\ref{CE7}) and a periodic path $r(w)$ of (\ref{CE8}). 
         }
\end{minipage} 
\end{figure}

First we re-express $E_{Cas}^{\La 0}$ using a simple identity : 
$l=\int_0^ldw$ ($w$: a regularization axis). 
\bea
E_{Cas}^{\La 0}/(2L)^2
=\frac{1}{2^2\pi^3}\int_0^ldw\int_{k\leq\La}P(k)2\pi k^2dk\nn
=\frac{1}{2^2\pi^3}\int_0^ldw (-1)\int_{r\geq\La^{-1}}P(1/r)(-1)2\pi r^{-4}dr.\nn
P(k)\equiv k,\q r\equiv \frac{1}{k}
,
\label{CE6}
\eea
where the integration variable changes from the momentum
($k$) 
to the coordinate 
($r\equiv 1/k$). The { integration region} in ($R,w$)-space is the infinite rectangular 
shown in Fig.\ref{2DregionI9}. 

The expression (\ref{CE6}) severely diverges as $\La\ra\infty$. 
In order to regularize it, we first replace the summation over all coordinate-space points
($R,w$) with the summation over all possible paths (path-integral). 
For each path we introduce the damping factor as in the same way in Sec.3.  Hence 
the above expression is replaced (regularized) by the following path-integral expression: 
\bea
{E_{Cas}^{\Wcal}}'/(2L)^2
=\frac{2\pi}{2^2\pi^3}\int_{\mbox{all paths $r(w)$}}\nn
\prod_w\Dcal r(w) \left[ \int dw'P(\frac{1}{r(w')}) r(w')^{-4}
                 \right]
\exp\left\{ -\Wcal[r(w)]\right\}
,
\label{CE7}
\eea
where the integral is over { all paths} $r(w)$ which are defined between $0\leq w\leq l$ 
and whose value is above $\La^{-1}$, 
as shown in Fig.\ref{2DpathI9}. 
$\Wcal[r(w)]$ is some {\it damping functional}. 
$\Wcal[r(w)]=0$ corresponds 
to (\ref{CE6}). The slightly-more-restrictive regularization is 
\bea
E_{Cas}^{\Wcal}/(2L)^2
=\frac{2\pi}{2^2\pi^3}\int_{\La^{-1}}^\infty d\rho\int_{r(0)=r(l)=\rho}\nn
\times\prod_w\Dcal r(w) \left[ \int dw'P(\frac{1}{r(w')}) r(w')^{-4}
                  \right]
\exp\left\{ -\Wcal[r(w)]\right\}~\geq 0
,
\label{CE8}
\eea
where the integral is over all {\it periodic} paths. 
Note that the above {\it regularization} keeps 
the { positive-definite} property. 
It is mainly defined by the choice of $\Wcal[r(w)]$. 
In order to specify it, we {\it introduce} the following metric in ($R,w$)-space\cite{SI1004}. 
\bea
\mbox{Dirac Type}:\ 
ds^2=dR^2+V(R)dw^2,\ V(R)=\Omega^2R^2
,
\label{CE9b}
\eea
or
\bea
\mbox{Standard Type}:\ 
ds^2=\frac{1}{dw^2}(dR^2+V(R)dw^2)^2,\nn 
                           V(R)=\Omega^2R^2
,
\label{CE9}
\eea
where $\Omega$ is the regularization constant. 
(When $V(R)=1$, $w$ is the familiar Euclidean time. ) 
On a path $R=r(w)$, the induced metric and the length $L$ 
is given as follows. As the { damping 
functional} $\Wcal[r(w)]$, we take the {\it length} $L$. 
\bea
ds^2=dw^2({r'}^2+\Omega^2r^2)^2,\q r'\equiv \frac{dr}{dw},\nn
L=\int ds=\int ({r'}^2+\Omega^2r^2)dw,\nn
\Wcal[r(w)]\equiv \frac{1}{2\al'}L
=\frac{1}{2\al'}\int ({r'}^2+\Omega^2r^2)dw
,
\label{CE10}
\eea
where $\al'$ and $\Omega$ are  the regularization parameters.  They can be regarded as 
the model parameters for the {\it statistical ensemble} which is taken as the regularization. 
The limit $\al'\ra\infty$ corresponds to (\ref{CE6}). 

Numerical calculation can evaluate $E_{Cas}^\Wcal$ (\ref{CE8}), and we expect the 
following form\cite{SI0801,SI0812,SI0909,SI1205}. 
\bea
\frac{E_{Cas}^\Wcal}{(2L)^2}=\frac{a}{l^3}(1-3b\ln~(l\La))
\com
\label{CE11}
\eea
where $a$ and $b$ are some constants. 
$a$ should be positive because of the positive-definiteness of (\ref{CE8}). 
The present regularization result has, like the ordinary 
renormalizable ones such as the coupling 
in QED, the {\it log-divergence}. 
The divergence can be renormalized into 
the {\it boundary} parameter $l$. 
This means $l$ {\it flows} according to the { renormalization group}. 
\bea
l'=l(1-3b\ln(l\La))^{-\frac{1}{3}},\nn
\be\equiv \frac{d\ln(l'/l)}{d\ln\La}=b,\q |b|\ll 1
,
\label{CE12}
\eea
where $\be$ is the renormalization group function, and we assume $|b|\ll 1$. 
The sign of $b$ determines whether the length separation increases ($b>0$) or decreases ($b<0$) 
as the measurement resolution becomes finer ($\La$ increases). In terms of the usual terminology, 
{\it attractive} case corresponds to $b>0$, and {\it repulsive} case to $b<0$. 
Compare the above way of determining the force-direction with that in Sec.\ref{CE} 
(ordinary case) where the relation $F=-\pl V/\pl x$ is necessarily used.

\section{
Conclusion
}
The electromagnetism in substance is formulated in the 
geometrical way. 
The dispersion relation is introduced by the 3 dim hyper-surface (\ref{geo3}) 
in ($\om,K^i$) space. The permittivity and the permeability are 
regarded as the metric defined on the hyper-surface. 
The micro fluctuation effect is taken into account by the generalized 
path-integral (\ref{geo10}). 
The new model parameter $\al'$ (string tension) is introduced, which is 
necessary in the present formulation using the geometrical quantity, {\it area} $A$. 
We point out the renormalization of the {\it boundary} parameters, 
$T$ (boundary of the frequency $\om$) and $H_0$ (4 dim curvature), 
takes place in the treatment of IR and UV divergences. 
In relation to the problem of the attractive or repulsive 
force, Lifshitz formula is explained in the context of the regularization of the 
quantum field theory. The regularization is basically the same as that in Sec.\ref{CE}, 
and there appears no renormalization of $l$. 
The new regularization is applied to Casimir energy calculation and 
compared with the ordinary case. The advantageous points are 1) the positivity is 
preserved in the regularization, 2) attractive or repulsive is determined by 
the sign of the {\it renormalization group} $\beta$-function for the boundary parameter $l$.


\begin{thebibliography}{15}
\bibitem{Malda9711}
J.M. Maldacena, Adv.Theor.Math.Phys.\textbf{2}(1998)231 [Int. J. Theor. Phys.\textbf{38}(1999)1113], 
arXiv:hep-th/9711200
\bibitem{GKP9802}
S. S. Gubser, I. R. Klebanov and A. M. Polyakov, \PL \textbf{B428}(1998)105, arXiv:hep-th/9802109
\bibitem{Witten9802}
E. Witten, Adv. Theor. Math. Phys.\textbf{2}(1998)253, arXiv:hep-th/9802150
\bibitem{SusWit9805}
L. Suskind and E. Witten, "The Holographic Bound in Anti-de Sitter Space", arXiv:hep-th/9805114
\bibitem{HenSken9806}
M. Henningson and K. Skenderis, JHEP \textbf{9807}(1998)023, arXiv:hep-th/9806087\\
M. Henningson and K. Skenderis, Fortsch.Phys. \textbf{48}(2000)125, arXiv:hep-th/9812032
\bibitem{Casimir48}
H,B,G. Casimir, Proc. Kon. Nederl. Akad. Wet. \textbf{51}(1948)793
\bibitem{ICSF2010}
S. Ichinose, J.Phys:Conf.Ser.\textbf{258}(2010)012003, arXiv:1010.5558, 
Proc. of Int. Conf. on Science of Friction 2010 
(Ise-Shima, Mie, Japan, 2010.9.13-18).
\bibitem{BMM01}
M. Bordag, U. Mohideen and V.M. Mostepanenko, Phys.Rep.\textbf{353}(2001)1
\bibitem{LL80}
E.M. Lifshitz and L.P. Pitaevskii, Statistical Physics Part2, Vol.9 of Course of 
Theoretical Physics, Butterworth-Heinemann, Oxford, UK
\bibitem{Tamm24}
I.E. Tamm, Zh. Rus. Fiz.-Khim. Obshchestva, Otd. Fiz. \textbf{56}, 248(1924)
\bibitem{Skrotskii57}
G.V. Skrotskii, Dokl. Akad. Nauk SSSR \textbf{114}, 73(1957) [Soviet Physics Doklady 2, 226(1957)]
\bibitem{Plebanski60}
J. Plebanski, Phys. Rev. \textbf{118}, 1396(1960)
\bibitem{GHWW09}
G.W. Gibbons, C.A.R. Herdeiro, C.M. Warnick and M.C. Werner, Phys.Rev.D \textbf{79},044022(2009)
\bibitem{SI0801}  
S. Ichinose, 
\PTP\textbf{121}(2009)727, ArXiv:0801.3064v8[hep-th]. 
\bibitem{KK06}
O. Kenneth and I. Klich, Phys.Rev.Lett.\textbf{97}(2006)160401
\bibitem{SI0812}  
S. Ichinose, "Casimir Energy of 5D Warped System and Sphere Lattice Regularization", 
ArXiv:0812.1263[hep-th], US-08-03, 61 pages. 
\bibitem{SI0909} 
S. Ichinose, 
J. Phys. :\ Conf.Ser.\textbf{222}(2010)012048. Proceedings of 
First Mediterranean Conference on Classical and Quantum Gravity 
(09.9.14-18, Kolymbari, Crete, Greece). ArXiv:1001.0222[hep-th] 
\bibitem{SI1205} 
S. Ichinose, 
"Casimir Energy of the Universe and the Dark Energy Problem", 
To appear in Proceedings of DSU2011 (2011.9.26-30, Beijin, China). ArXiv:1205.1316[hep-th] 
\bibitem{SI1004}  
S. Ichinose, "Geometric Approach to Quantum Statistical Mechanics and Minimal Area Principle", 
ArXiv:1004.2573[hep-th], US-10-03, 28 pages. 
\end{thebibliography}
\end{document}